\documentclass[11pt]{article}
\usepackage[a4paper, left=20mm, right=20mm, top=20mm, bottom=40mm]{geometry}
\usepackage{bm,amssymb,amsfonts,amsmath} 
\usepackage{graphicx} 

\newcommand{\AM}{\mathcal{H}_\mathrm{AM}}
\newcommand{\BM}{\mathcal{H}_\mathrm{BM}}
\newcommand{\FS}{\mathcal{H}_\mathrm{FS}}
\newcommand{\cond}{\,|\,}
\newcommand{\bmD}{\bm{D}}
\newcommand{\bmw}{\bm{w}}

\newcommand{\hmodel}{\mathcal{H}}
\newcommand{\info}{\mathcal{I}}
\newcommand{\logev}{\mathrm{LE}}
\newcommand{\matC}{\mathbb{C}}
\newcommand{\matH}{\mathbb{H}}
\newcommand{\deta}{\ensuremath{\eta_{\scriptscriptstyle \Delta}}}
\newcommand{\dphi}{\ensuremath{\phi_{\scriptscriptstyle \Delta}}}
\begin{document}

\begin{center}
  \textbf{\large Bayesian model comparison for one-dimensional azimuthal
    correlations}\\[6pt]
    \textbf{in 200GeV AuAu collisions}\footnote{%
    Proceedings of the XLV International Symposium on Multiparticle Dynamics, 
    Wildbad Kreuth, 5--9 October 2015}\\[16pt]
  Hans C.\ Eggers$^{a,b}$ Michiel B. de Kock$^a$ and Thomas A.\ Trainor$^c$\\[12pt]
  $^a$\textit{Department of Physics, Stellenbosch University, 7600
  Stellenbosch, South Africa} \\
  $^b$\textit{National Institute for Theoretical Physics (NITheP), 7600
  Stellenbosch, South Africa}\\
  $^c$\textit{CENPA 354290, University of Washington, Seattle, Washington 98195,
  United States}
\end{center}

\abstract{%
  In the context of data modeling and comparisons between different
  fit models, Bayesian analysis calls that model best which has the
  largest evidence, the prior-weighted integral over model parameters
  of the likelihood function. Evidence calculations automatically take
  into account both the usual chi-squared measure and an Occam factor
  which quantifies the price for adding extra parameters. Applying
  Bayesian analysis to projections onto azimuth of 2D angular
  correlations from 200 GeV AuAu collisions, we consider typical model
  choices including Fourier series and a Gaussian plus combinations of
  individual cosine components. We find that models including a
  Gaussian component are consistently preferred over pure
  Fourier-series parametrizations, sometimes strongly so. %
  For 0--5\% central collisions the Gaussian-plus-dipole model
  performs better than Fourier Series models or any other combination
  of Gaussian-plus-multipoles.
} %

\section{Introduction}
\label{sec:intro}

A significant problem has emerged in the interpretation of high-energy
collision data from the relativistic heavy ion collider (RHIC) and the
large hadron collider (LHC). Two classes of data models with divergent
physics implications are invoked to support competing narratives.  One
narrative interprets results with concepts from high-energy physics in
which the essential phenomenon is dijet production; the other uses 
quark-gluon plasma and collective flow concepts in which the essential
phenomenon is a flowing, dense medium while dijets play only a
subordinate role.

To place evaluation and comparison of competing models on a scientific
footing we require a formally consistent and statistically sound
framework. We suggest that Bayesian inference provides the formal
consistency and the necessary mathematical framework for representing
prior knowledge, evaluating candidate models for different parameter
values and thereby establishing value judgments on models as a
whole. In this contribution, we show by example how the Bayesian
framework successfully evaluates and compares models for that type of
one-dimensional correlation data which has played a large role in
claimed support for these narratives. The exercise also serves to
pinpoint the weaknesses of the conventional minimum-$\chi^2$ and
maximum-likelihood approaches.

\section{Data and parametrization}
\label{sec:dpm}

The data analyzed below were published by the STAR Collaboration
\cite{anomalous} in the form of two-dimensional binned histograms
derived from 1.2 million 200GeV AuAu collision events in eleven
centrality classes. The relevant autocorrelations
\begin{equation}
  A(\dphi,\deta) = \Delta\rho/\sqrt{\rho_{\rm ref}} =
  \rho_0 \left[(\rho/\rho_{\rm ref}) - 1 \right]
\end{equation}
are measured on azimuthal and pseudorapidity difference variables
$\dphi = \phi_1 - \phi_2$ and $\deta = \eta_1 - \eta_2$, pair
densities $\rho(\dphi,\deta)$ and $\rho_{\rm ref}(\dphi,\deta)$ and
the one-particle differential cross section, $\rho_0 = d^2n_{ch}
/d\eta d\phi$.

Here, we consider one-dimensional projections of the 2D
autocorrelations onto azimuth difference $\dphi$ with 24 bins of size
$\pi/12$.  As the data are symmetrized on both $\deta$ and $\dphi$,
there are only 13 unique points. These are reduced to $N=11$ data
points by excluding from fits the $\dphi=0$ bin containing the
Bose-Einstein peak and constraining the 1D projections to sum to
zero.

Since the 11-parameter 2D parametrization (or data model) described in
Ref.~\cite{anomalous} describes the corresponding 2D data rather well,
we adapted it to the present 1D projection. Taking into account the
zero-sum constraint, this results in what we call the Basic Model;
abbreviating $\phi \equiv \dphi$,
\begin{align}
  \label{parbm}
  f(\phi,\bmw\cond\BM) &=
  A_{\rm 1D} \left\{  \exp\left[- \frac{1}{2} 
      \left( \frac{\phi}{ \sigma_{\phi}} \right)^2 \right] %
    - \frac{\sigma_{\phi}}{\sqrt{2\pi}} \right\} 
  - A_{\rm D}'\, 2\cos(\phi)
\end{align}
with three parameters $\bmw = ( A_{1D},\sigma_{\phi}, A_{\rm D}')$
reflecting the basic same-side (SS) peak and away-side (AS) dipole
structure. The Basic Model $\BM$ is supplemented in Augmented Models
$\AM$ by a quadrupole term $- A_{\rm Q}'\, 2\cos(2\phi)$, a sextupole
$- A_{\rm S}'\, 2\cos(3\phi)$, and an octupole $- A_{\rm O}'\,
2\cos(4\phi)$ in various combinations.
The factor 2 entering the cosine terms is designed to reflect a
corresponding Fourier Series (FS) parametrization popular within
the collective-flow paradigm \cite{2004},
\begin{align}
  \label{parfs}
  f(\phi,\bmw\cond\FS) &= 2\sum_{j=1}^K w_j \cos(j \phi).
\end{align}
In our analysis, the number of free parameters $K$ varies from 1 to 11
in the FS case, while in the Basic and Augmented Models $K$ varies
from 3 to 6.

\section{Model comparison}

The two goals of fitting a parametrization to data are model
comparison and, given a particular model $\mathcal{H}$, finding the
best-fit values of its parameters. In the present case, the data $\bmD
= (y_1,\ldots,y_N)$ are given by binwise autocorrelations $A(\phi)$ at
mid-bin sample points $\phi_n$ while the competing models include
$\BM$, $\AM$ and $\FS$ with various values of $K$.

In the usual \textbf{frequentist approach}, model comparison is
effected by searching the space of parameters $\bmw$ to find the
global minimum $\chi_0^2$ of
\begin{align}
  \chi^2 = \sum_{n=1}^N \left( \frac{y_n - f(\phi_n,\bmw\cond\hmodel)}{\sigma_n}\right)^2
\end{align}
with $\sigma_n$ the experimental standard errors. Model $\hmodel_1$
with $K_1$ parameters is deemed to be better than Model $\hmodel_2$
with $K_2$ parameters if $\min(\chi^2_1)/(N-K_1) <
\min(\chi^2_2)/(N-K_2)$.
Underlying the $\chi^2$ criterion is the assumption of independent
Gaussian fluctuations of deviations between data and parametrization,
\begin{align}
  y_n - f(\phi_n,\bmw\cond\hmodel) \ &=\ \varepsilon_n 
  \qquad\text{with}\qquad 
  p(\varepsilon_n) \ =\ \text{ Gaussian with standard deviation } \sigma_n
\end{align}
which results in a likelihood
\begin{align}
  \label{lkh}
  p(\bmD\cond \bmw,\hmodel) &= \frac{e^{-\chi^2/2}}{(2\pi)^{N/2}\prod_n\sigma_n}.
\end{align}
When parametrizations $f(\phi,\bmw\cond\hmodel)$ are linear in $\bmw$,
the Hessian
\begin{align}
  \matH_{k\ell} = \frac{1}{2}\,
  \frac{\partial^2\chi^2}{\partial w_k\,\partial w_\ell} 
\end{align}
is $\bmw$-independent and the likelihood takes on a
Gaussian form in parameter space 
around the mode $\tilde{\bmw}$ in terms of $\matH$ or equivalently the
covariance matrix $\matC = \matH^{-1}$.

\textbf{Bayesian model comparison} utilizes the same likelihood
(\ref{lkh}) and hence also $\chi^2$ and the covariance matrix. Crucially,
however, Bayesian model comparison is based not on the maximum
likelihood or $\chi_0^2$ but on the so-called \textit{evidence} for
data $\bmD$ within model $\hmodel$,
\begin{align}
  \label{pvv}
  p(\bmD\cond\hmodel) &= \int
  d\bmw\,p(\bmD\cond\bmw,\hmodel)\,p(\bmw\cond\hmodel),
\end{align}
which averages the likelihood over all values of $\bmw$, weighted by
the \textit{prior} $p(\bmw\cond\hmodel)$ for the parameters.

The reason why evidence is a good measure for model comparison lies
within Bayes' Theorem itself. Starting with the joint data-parameter
probability 
$ p(\bmD,\bmw\cond\hmodel) %
= p(\bmD\cond\bmw,\hmodel) \, p(\bmw\cond\hmodel)%
\ =\ p(\bmw\cond\bmD,\hmodel) \, p(\bmD\cond\hmodel), $
Bayes' Theorem states that, once data $\bmD$ is known, the posterior
probability for parameters $\bmw$ can be found from the likelihood,
prior and evidence by
\begin{align}
  p(\bmw\cond\bmD,\hmodel)
  &= \frac{p(\bmD\cond\bmw,\hmodel) \, p(\bmw\cond\hmodel)}{p(\bmD\cond\hmodel)}
  \qquad\qquad
  \text{(posterior)} %
  = \frac{\text{(likelihood)}\,\text{(prior)}}
  {\text{(evidence)}}. 
\end{align}
Similarly on a higher level $ p(\bmD,\hmodel) = p(\bmD\cond\hmodel) \,
p(\hmodel) \ =\ p(\hmodel\cond\bmD) \, p(\bmD) $ implies that, given
data $\bmD$, the probability for the entire model $\hmodel$ including
all possible parameter values is
\begin{align}
  p(\hmodel\cond\bmD)
  &= \frac{p(\bmD\cond\hmodel) \, p(\hmodel)}{p(\bmD)}
\end{align}
which, used twice for competing models $\hmodel_1$ and $\hmodel_2$,
says that model comparison given data $\bmD$ can be expressed 
as a ratio of evidences,
\begin{align}
  \label{hmc}
  \frac{p(\hmodel_1| \bmD)} {p(\hmodel_2|\bmD)}
  &=
  \frac{p(\bmD|\hmodel_1)} {p(\bmD|\hmodel_2)}
  \;
  \frac{p(\hmodel_1)} {p(\hmodel_2)}
  \ =\
  \frac{p(\bmD|\hmodel_1)} {p(\bmD|\hmodel_2)}
  \ =\ 
  \frac{\text{evidence for } \hmodel_1}
  {\text{evidence for } \hmodel_2}
\end{align}
on setting equal \textit{a priori} probabilities for models,
$p(\hmodel_1) = p(\hmodel_2) = 1/2$.

The choice of \textbf{parameter prior} $p(\bmw\cond\hmodel)$ is
governed by available information or lack thereof. We use uniform
priors, implying indifference for any value of $\bmw$ within the
bounds set by cutoffs or prior widths $\Delta_k$,
\begin{align} 
  \label{priordelt}
  p(\bmw|\hmodel) &= \textstyle\prod_{k=1}^K \Delta_k^{-1}.
\end{align}
We assign the same widths across different models for corresponding
quantities. In particular, we set $\Delta_k = 1/3$ for all
$k=1,\ldots,K$ cosine coefficients, based on knowledge of typical
coefficient magnitude. The interval for $\sigma_{\phi}$ is set
to $[0,\pi/2]$ in accordance with its definition as a same-side peak
width. To the prior for the Gaussian amplitude parameter $A_{\rm D}'$
we assign width $\Delta = 1$ since correlation-structure amplitudes
such as peak-to-peak excursions are generally $< O(1)$.

For uniform priors, the evidence integrals (\ref{pvv}) can be carried
out analytically in terms of the \textbf{Laplace approximation} which
neglects third- and higher-order terms in the Taylor expansion of the
log likelihood and is exact for linear systems. In our case, the log
likelihood is nonlinear only in the parameter $\sigma_\phi$. The
Laplace Approximation also requires that the effect of finite
integration limits set by the $\Delta_k$ be negligible. Within these
constraints, the evidence takes on a simple form in terms of the
maximum likelihood, prior and covariance matrix,
\begin{align}
  p(\bmD|\hmodel) %
  &= p(\bmD|\tilde{\bmw},\hmodel)
  \; p(\tilde{\bmw}\cond\hmodel)\, (2\pi)^{K/2} \sqrt{\det \matC}\,.
\end{align}
Since $\chi_0^2$ is related to the maximum likelihood by
$  \chi_0^2 = -2 \log p(\bmD\cond\tilde{\bmw},\hmodel)$,
we can write the corresponding log evidence in terms of $\chi_0^2$ and
an \textit{information} $\info$,
\begin{align}
  \label{nfi}
  -2\logev &= \chi_0^2 + 2 \info, \\
  \label{nfj}
  \info \;&= 
  -\log\Bigl[ p(\tilde{\bmw}|\hmodel)\,(2\pi)^{K/2}\sqrt{\det\matC}\,\Bigr].
\end{align}
The smaller $-2\logev$ for a given model, the larger its evidence.  As
shown in Eq.~(\ref{hmc}), $-2\logev$ for different models have no
absolute meaning but must always be interpreted in terms of evidence
ratios (also called \textit{odds} in the literature) or of log
differences.

\begin{center}
\includegraphics[width=100mm,clip]{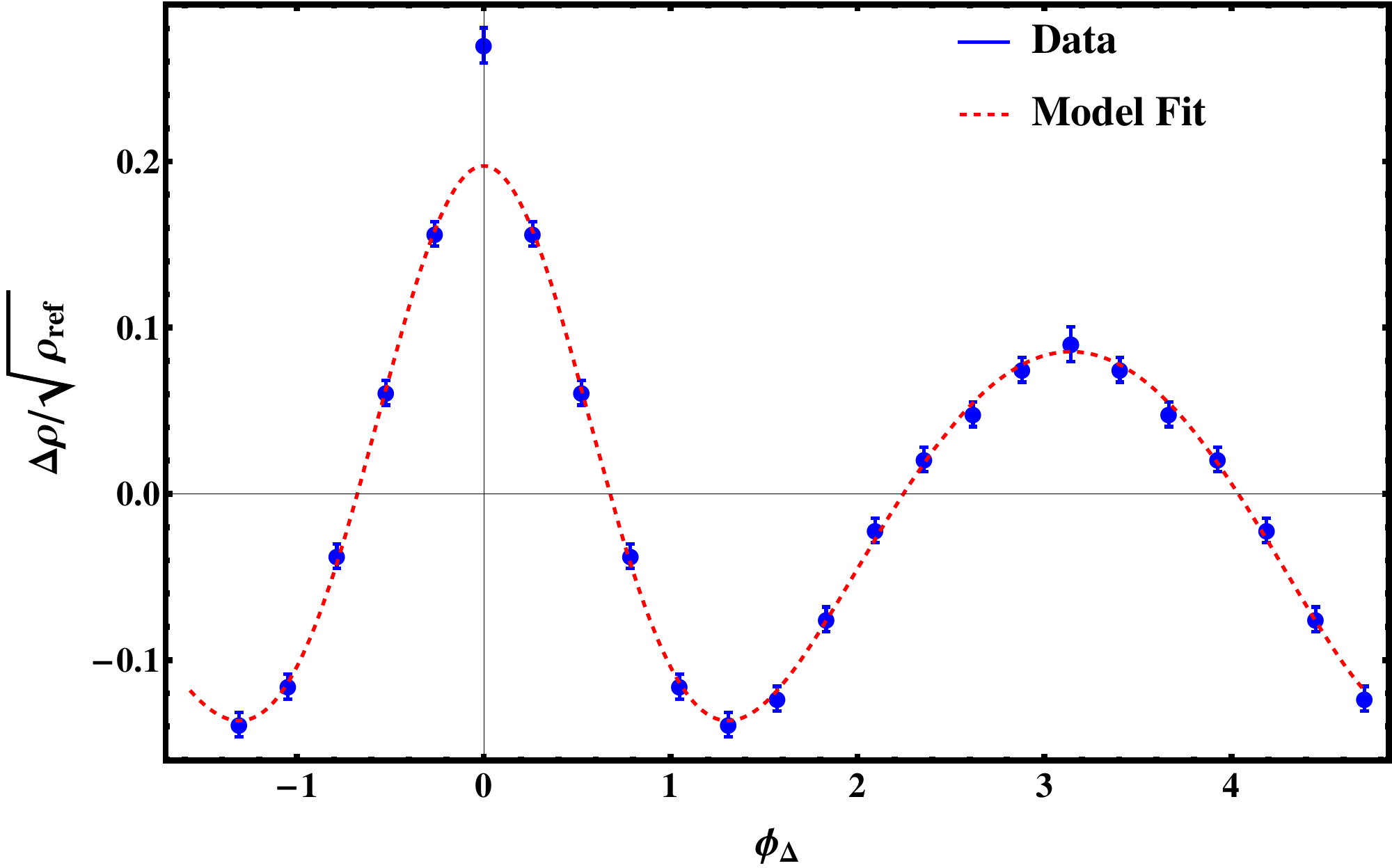}
\end{center}
\begin{quote}
  Figure 1. (Color online) Autocorrelation data projected onto one
  dimension for the most central events in 200 GeV AuAu collisions
  measured by STAR \cite{anomalous} along with the Basic Model best
  fit. Corresponding Fourier-Series fits with four or more terms
  appear identical on the scale of the figure. Eyeball viewing is not
  a reliable basis for model comparison. 
\end{quote}

\begin{center}
\includegraphics[width=100mm,clip]{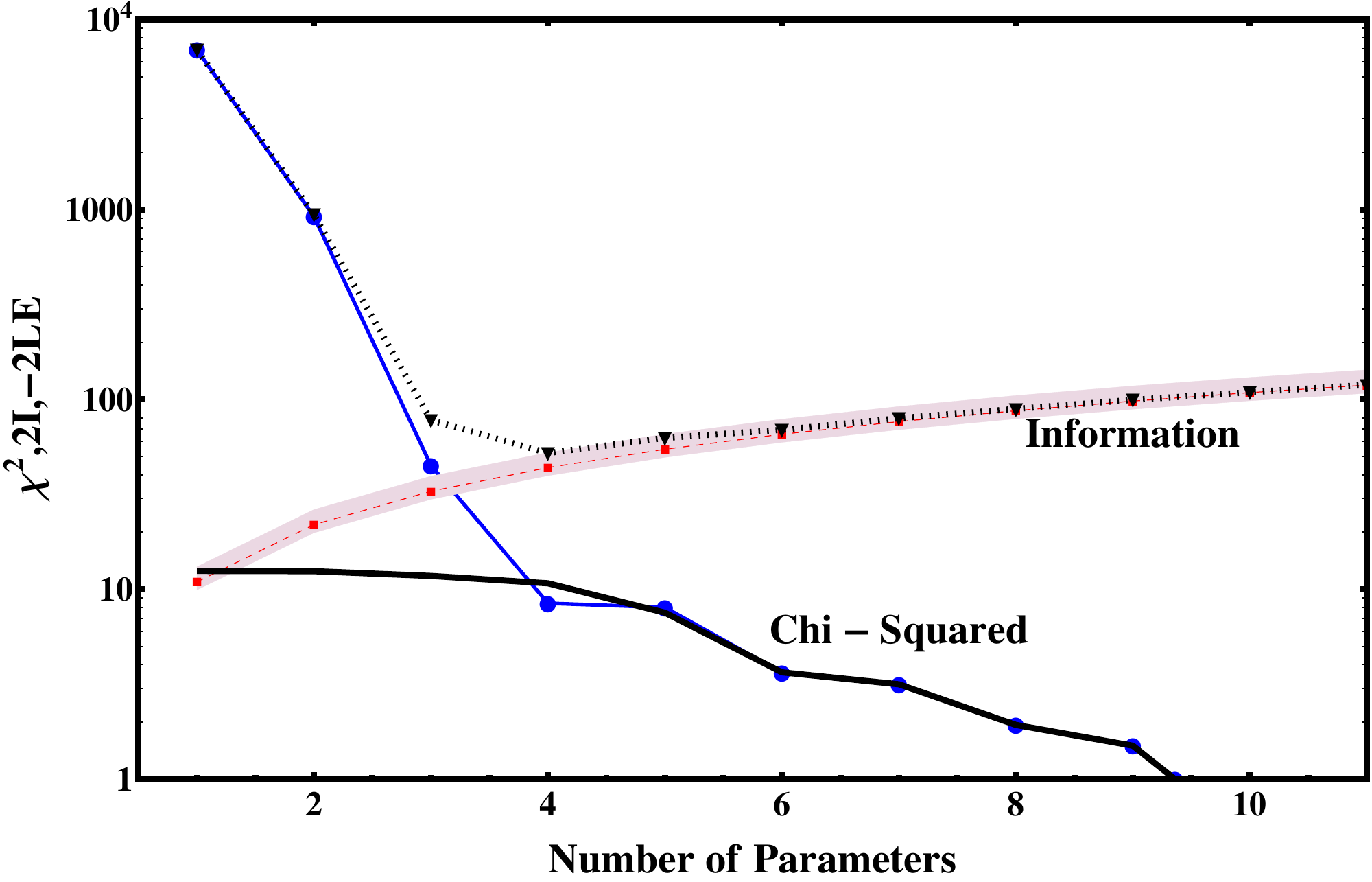}
\end{center}
\begin{quote}
  Figure 2. (Color online) Soid blue line: the usual minimum $\chi_0^2$,
  which becomes identical with the $\chi_0^2$ of the noise for $K\geq
  5$ shown as the black solid line. Red dashed line and band:
  Information $2\info$ plus possible uncertainty due to the choice of
  prior. The negative log-evidence $-2\logev$, the sum of $\chi_0^2$
  and $2\info$, is shown as the dashed black line. It has a clear
  minimum at $K=4$.
\end{quote}

\section{Results}

In this conference summary, we can show results only for the STAR
``Bin-10'' data covering the most-central 0--5\% collision events;
further results for Bin-0 peripheral collisions (83--94\%) and Bin-8
intermediate collisions (9--18\%) are presented and discussed in
Ref.~\cite{DeKock2015a}.

In Figure 1, 
the projected 1D data along with the
best-evidence fit for the Basic Model (\ref{parbm}) is
shown. Experimental error bars were scaled up by a factor $2$ to make
them visible in the plot.  As stated, the point at $\phi=0$ was
excluded since it contains the Bose-Einstein peak and charge-neutral
electron pairs from photoconversions.
The best-fit parameter values are $A_{1D} =0.57\pm 0.007$, $A'_D =
0.115\pm 0.002$ and $\sigma_{\phi} =0.635\pm 0.007$ while $\chi_0^2 =
12.5$ for $11-3$ degrees of freedom.
A fit of the Fourier Series model (\ref{parfs}) with $K\geq 4$ is
indistinguishable to the eye on the scale of the plot.  That does not
mean that the BM and FS models have the same $-2\logev$.

Figure 2, 
illustrates for the example of Fourier-Series models
why model comparison based on minimum $\chi^2$ alone is misleading
when the number of data points is small, as is the case here since
$N=11$. Plotted are various quantities as a function of the number of
free parameters $K$ in Eq.~(\ref{parfs}).  The solid blue line traces
the evolution of $\chi_0^2$ as more cosine terms are added.  As
expected, there is a dramatic decrease in $\chi_0^2$ up to $K=4$ but
little benefit for adding further terms beyond that as $K\geq 5$
Fourier Series models increasingly fit the noise represented by the
lower solid curve showing fits to the residuals (data minus Basic
Model).
The naive criterion of minimum $\chi_0^2$ is oblivous to this effect:
$\chi_0^2$ keeps decreasing and purely $\chi_0^2$-based model
comparison would hence select $K=9$ or higher. Note that the
equivalent plot for $\chi_0^2/(N-K)$ (not shown) barely changes the
general decrease with $K$.

The contribution of information $2\info$ of Eq.~(\ref{nfj}) is shown
as the red dashed curve. The shaded band shows the sensitivity to the
$\Delta_k=1/3$ prior width, using $\Delta_k=1$ and $\Delta_k = 1/5$ as
extremes. The resulting log-evidence $-2\logev$ of Eq.~(\ref{nfi}),
shown as the black dashed curve, has a clear minimum at $K=4$,
confirming the above conclusion that FS models with $K\geq 5$ are not
preferred.

\begin{center}
\includegraphics[width=100mm,clip]{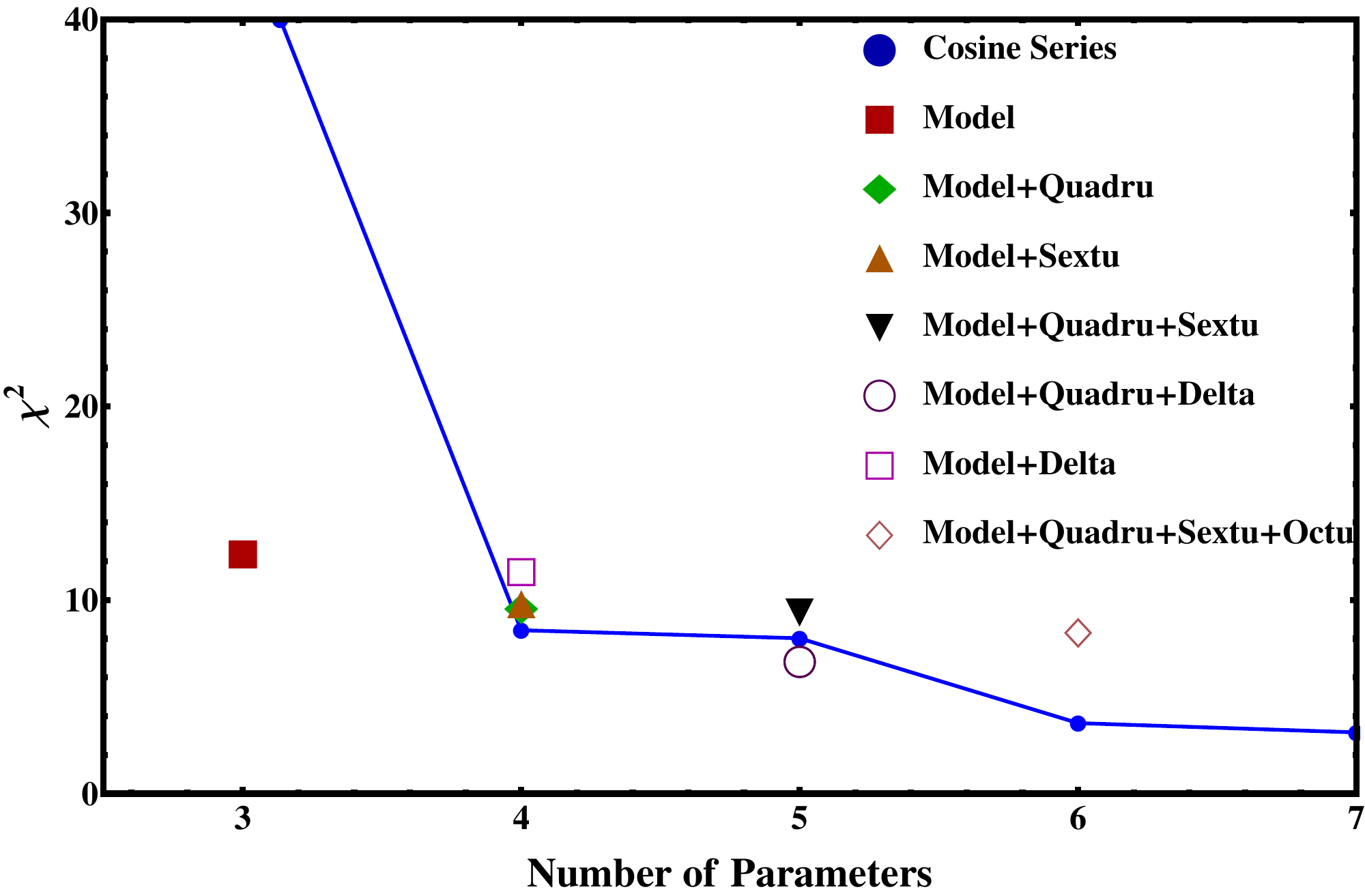}
\end{center}
\begin{quote}
  Figure 3(a). (Color online) Minimum-$\chi^2$ comparison of FS models
  (solid blue line) with Gaussian plus various combinations of
  multipoles. 
\end{quote}


\begin{center}
\includegraphics[width=100mm,clip]{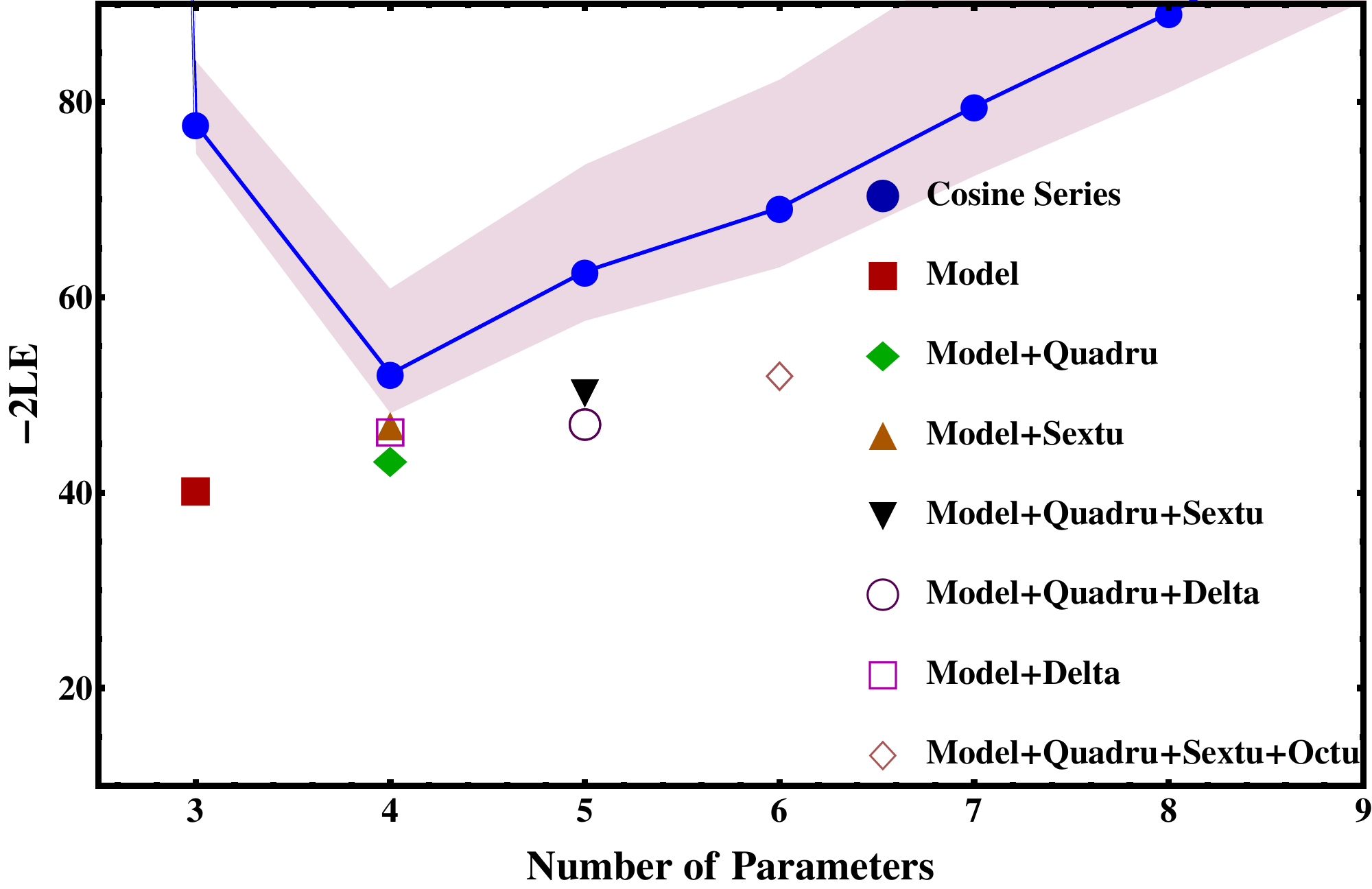}
\end{center}
\begin{quote}
  Figure 3(b). (Color online) The same models assessed with the
  Bayesian log-evidence criterion. The 3-parameter Basic Model of
  Eq.~(\ref{parbm}) (filled square) is the clear winner. Note that
  $-2\logev$ is a logarithmic scale.
\end{quote}

\noindent
In Figure 3(a), 
the scope of comparison between pure
$\chi_0^2$ and evidence-based model comparison is widened to include
FS, Basic Model and Augmented Models with various combinations of
higher-order multipoles. Consideration of $\chi_0^2$ alone as in
Fig.~3(a) 
would make the $K\geq 6$ FS models the winner. By
contrast, the Occam penalty of $2\info$ forming part of $-2\logev$ in
Figure 3(b) 
effects a dramatic re-ordering: the 3-parameter
Basic Model has substantially smaller $-2\logev$ (larger evidence)
than all other combinations. In addition, FS models are universally
rejected in favour of Gaussian-plus-multipole models with the same
number of parameters.

\section{Conclusions}

\textbf{Physics interpretation:} As shown, the evidence for a simple
same-side Gaussian plus away-side dipole is larger than for any Fourier
Series-based model. While our approach and results must naturally be
tested and confirmed with other data, measurement quantities and
models, our results do cast serious doubt on the interpretation of
azimuthal correlations as indicators of collective flow.\\

\noindent
\textbf{Lessons for data analysis:} Physics conclusions must rely on
the integrity of the underlying data analysis methods and assumptions.
Below, we provide in tabular form a motivation why Bayesian
evidence-based model comparison is superior to simple $\chi^2$
fitting. The sometimes very different conclusions shown above suggest
that any physical interpretation based on simple $\chi^2$ fitting of
one-dimensional correlation data may be debatable or outright
wrong. We secondly emphasize that no model or ``goodness of fit'' has
any meaning unless it is compared to another competing model.
\\[8pt]
\renewcommand{\arraystretch}{1.15}
\begin{tabular}{|p{66mm}|p{70mm}|}
    \hline %
    \textbf{Traditional use of $\chi^2\rule[0pt]{-4pt}{12pt}$} 
    & \textbf{Bayesian Inference} \\ %
    \hline %
    \textbf{Parameter estimation:}
    Find maximum likelihoods 
    $\max p(\bm{D}\cond\bmw_1,\hmodel)$ and $\max p(\bm{D}\cond\bmw_2,\hmodel_2)$ 
    which determine \textbf{single best fit parameters} 
    $\tilde{\bmw}_1, \tilde{\bmw}_2$.  This is equivalent to
    &  \textbf{Parameter estimation:} Find evidences
    $p(\bm{D}\cond \hmodel_1)$ and $p(\bm{D}\cond \hmodel_2)$
    which consider \textbf{all possible parameter values}. 
    This is equivalent to \\
    $\qquad\min\chi_1^2 = -2\ln p(\bm{D}\cond\tilde{\bmw}_1,\hmodel_1)$
    & $\qquad -2\,\logev_1 = -2\ln p(\bm{D}\cond\hmodel_1)$ \\[2pt]
    $\qquad \min\chi_2^2 = -2\ln p(\bm{D}\cond\tilde{\bmw}_2,\hmodel_2)$ 
    & $\qquad -2\,\logev_2 =  -2\ln p(\bm{D}\cond\hmodel_2)$ \\[2pt]
    \hline
    \textbf{Estimating uncertainties:}
    $\sigma(\bmw)$ is estimated from parameter covariance matrix $\matC$. %
    & \textbf{Estimating uncertainties:} Posterior $p(\bmw\cond\bm{D},\hmodel)$ %
    provides comprehensive information on parameters, including $\sigma(\bmw)$.
    \\
    \hline
    \textbf{Model comparison:} & \textbf{Model comparison:} \\
    Comparison of $\quad \chi_1^2/(N-K_1) \quad\text{to}\quad 
    \chi_2^2(N-K_2)$ 
    & Comparison of
    $\quad -2\,\logev_1\quad \text{to}\quad -2\,\logev_2$  \\[0pt] %
    uses only maximum likelihood $\displaystyle p(\bm{D}\cond\tilde{\bmw},\hmodel)$  %
    and ignores covariances of $\bmw$.
    & uses covariances of $\bmw$ through $\matC$ since $\qquad\qquad$
    $\displaystyle p(\bm{D}\cond\hmodel) = 
    \displaystyle p(\bm{D}\cond\tilde{\bmw},\hmodel)
    \; p(\tilde{\bmw}\cond\hmodel)\, (2\pi)^{K/2} \sqrt{\det \matC}$\,.
    \\
    \hline
    \textbf{Model construction:} & \textbf{Model construction:} \\
    Explore different likelihoods &
    Explore different priors and likelihoods\\
    \hline
    Accurate only for large $N$ & Accurate for all $N$ \\
    \hline %
  \end{tabular}

\vspace*{16pt}

\noindent
\textbf{Acknowledgements:}
  HCE thanks the Alexander von Humboldt Foundation and the
  \textit{Excellence Cluster Universe} for support. This work was
  supported in part by a Stellenbosch Consolidoc fellowship, the
  National Institute for Theoretical Physics and the National Research
  Foundation of South Africa.



\end{document}